\documentstyle[12pt,amsmath,epsfig,psfrag]{article}

\newcommand{\f}{\frac}
\newcommand{\non}{\nonumber \\}
\newcommand {\beq}{\begin{equation}}
\newcommand {\eeq}{\end{equation}}
\newcommand {\beqa}{\begin{eqnarray}}
\newcommand {\beqal}{\begin{eqnarray}\label}
\newcommand {\eeqa}{\end{eqnarray}}
\newcommand {\bc}{\begin{center}}
\newcommand {\ec}{\end{center}}
\newcommand {\s}{\sigma}
\newcommand {\al}{\alpha^{'}}
\newcommand {\als}{\alpha^{'2}}
\newcommand {\alp}{\alpha}
\newcommand {\th}{\theta}
\newcommand {\vth}{\vartheta}

\newcommand {\pa}{\partial}
\newcommand {\de}{\delta}
\newcommand {\De}{\Delta}
\newcommand {\ep}{\epsilon}

\newcommand {\kpe}{k_{\perp}}
\newcommand {\Tr}{\mbox{Tr}}

\newcommand{\expt}[1]{\left\langle #1 \right\rangle}
\newcommand {\ps}{\psi}
\newcommand {\bps}{\bar{\psi}}
\newcommand{\alpb}
{\left[\begin{smallmatrix}
\alp \\ \beta 
\end{smallmatrix}\right]}
\def\nn{\nonumber}
\newcommand{\oo}
{\left[\begin{smallmatrix}
1/2 \\ 1/2
\end{smallmatrix}\right]}
\def\nn{\nonumber}
\newcommand{\zo}
{\left[\begin{smallmatrix}
0\\ 1/2
\end{smallmatrix}\right]}
\def\nn{\nonumber}
\newcommand{\oz}
{\left[\begin{smallmatrix}
1/2 \\ 0
\end{smallmatrix}\right]}
\newcommand{\zz}
{\left[\begin{smallmatrix}
0 \\ 0 
\end{smallmatrix}\right]}

\begin{document}

\title{
\hfill\parbox{4cm}{\normalsize IMSC/2005/07/18}\\
\vspace{1cm}
Aspects of Open-Closed Duality in a Background B-Field II
\author{S. Sarkar and B. Sathiapalan \footnote{email:
\{swarnen, bala\}@imsc.res.in}\\
\\
{\em Institute of
Mathematical Sciences}\\
{\em Taramani}\\{\em Chennai, India 600113}}}
\maketitle

\begin{abstract}
\noindent
It was shown in [hep-th/0503009], in the context of bosonic theory that 
the IR singular terms that arise as a result of integrating out high 
momentum modes in nonplanar diagrams of noncommutative gauge theory can 
be recovered from low lying tree-level closed string exchanges. This 
follows as a natural consequence of world-sheet open-closed string 
duality. Here using the same setup we study the phenomenon for 
noncommutative ${\cal N}=2$ gauge theory realised on a $D_3$  
fractional brane localised at the fixed point of $C^2/Z_2$. The IR 
singularities from the massless closed string exchanges are exactly 
equal to those coming from one-loop gauge theory. This is as a result of 
cancellation of all contributions from the massive modes.
\end{abstract}
\newpage

\section{Introduction}

World sheet duality  
is a very
fundamental feature of string theories and one expects that because
of this there is a duality between open string theories and closed
string theories. AdS/CFT \cite{adscft} can be viewed as a particular 
realization of this
where to leading order we see a relation between the massless sectors
of the two theories. It is useful to study such dualities in other 
backgrounds
to further elucidate the key ingredients. One such background is
the constant $B$-field. As will be shown below because of the
regulatory nature of the $B$-field duality statements in some cases
can be made more sharply.

Open string dynamics in constant background $B$-field have been studied 
for the past few years with renewed interest for other reasons also.
 This is mainly due to the 
fact that the low energy dynamics of open strings in background constant 
$B$-field can be studied as a gauge theory on noncommutative spacetimes
 \cite{douglas1,schomerus,chu,Seiberg}.
Field theories on noncommutative spaces are interesting in themselves, 
however their embedding in string theory helps us to analyse them from a 
wider perspective. 

In the recent years there has been an extensive study of noncommutative 
field theories from two directions \cite{douglas}. One by starting from 
noncommutativity at the field theory level and exploring various 
phenomena that arise which may be absent in commutative models. The 
other direction is by studying them from the point of view of string 
theory. Most often the latter perspective has led to better 
understanding of the various unusual features in the 
noncommutative models.

It is now well known that a generic feature of noncommutative field 
theories is the appearance of infrared singularities by integrating out 
high momentum modes propagating in the loop, popularly known as UV/IR 
mixing. Various attempts have been 
made to interpret these results in the usual Wilsonian renormalisation 
group picture. It is shown that this phenomenon in noncommutative 
field theories fits into the usual notion of Wilsonian 
renormalisation group if we include an infrared cutoff \cite{myuvir}.
A different approach to cure the problem of UV/IR mixing 
has also been pursued \cite{sachin}. It is however,
not clear whether this is to be viewed as an inequivalent
quantisation and therefore a different theory or a different cure
to the infrared divergences.

However, the coupling of the ultraviolet to the infrared
is inherent in string theory and manifests itself as a consequence of 
open-closed string duality. With this hindsight it was proposed that 
the new IR singularities should appear at the field theory level by 
integrating out additional massless closed string modes that couple to 
the gauge theory \cite{minwalla,raamsdonk}. This was studied for the 
one loop $N$-point tachyonic amplitude in bosonic theory in 
\cite{Chaudhuri}. Interesting connections between the closed string 
tachyons and noncommutative divergences was shown in \cite{armoni}. 
Usually the ultraviolet divergences of the open string modes can be 
interpreted as infrared divergences from massless closed string exchanges.
In the presence of the background $B$-field these divergences are 
regulated and thus a quantitative analysis can be made. 
The one-loop two point diagram for open strings is a cylinder with 
a modular parameter $t$ and vertex operator insertions at the 
boundaries. The two point one-loop noncommutative field theory diagram 
results in the Seiberg-Witten limit by keeping surviving terms in the 
integrand for the integral over $t$ for $t \rightarrow \infty$. This 
limit suppresses all contributions from massive modes in the loop. The 
resulting diagram is that of the gauge theory with massless propagating 
modes. This amplitude is usually divergent in the ultraviolet when 
integrated over $t$. The 
source of ultraviolet divergence is the same as that of those in string 
theory i.e. $t\rightarrow 0$. It is therefore natural to analyse the 
amplitude directly in this limit when only the low lying closed string 
exchanges contribute. We have analysed this in \cite{my}.
We showed, in the context of bosonic 
string theory, that using 
open-closed string duality, one can recover the infrared divergent 
behaviour coming from nonplanar loop amplitudes in gauge theory from 
lowest lying tachyonic and massless tree-level closed string exchanges. 
However the correspondence could not be made exact as contributions from 
massive states in the bosonic theory are also relevant. Though the 
nature of the singularities matches with appropriate tensor structures, 
the coefficients are not equal. We argued 
that this equality can exactly occur in some special supersymmetric 
models. Open strings 
on fractional branes localised at the fixed point of $C^2/Z_2$ naturally 
satisfy these conditions. In this paper we study this issue in the setup 
of \cite{my} in the orbifold background.   

The fact that the ultraviolet behaviour of the one loop gauge theory is 
same as 
that of the massless closed string tree-level exchanges in this model
with ${\cal N}=2$ supersymmetry have been pointed out in 
\cite{douglas2}.
Further studies
as the consequence of this duality have been done by various authors
\cite{lerda1,Polchinski:2000mx,lerda2,Bertolini:2001gq}.
From the point of view 
of open-closed string duality this was analysed in \cite{DiVecchia}. 
See \cite{DiVecchia:2005vm} for a recent review and references therein. 
We show that the 
UV/IR mixing phenomenon of 
${\cal N}=2$ gauge theory can be naturally interpreted as a consequence 
of open-closed string duality in the presence of background $B$-field.
The effective action for the full two point 
function from gauge theory differs from that with closed string 
exchanges only by finite derivative corrections. 
However as far as the divergent UV/IR mixing term is concerned, it 
is exactly equal to the infrared contribution from the massless closed
string exchanges. Using world-sheet duality these modes can be 
identified as coming from the twisted NS-NS and R-R sectors. This 
model was also studied in \cite{rajaraman} in the context of closed 
string 
realization of the IR singular terms in gauge theory. Here 
in this paper we show how only the twisted NS-NS and R-R 
sectors closed strings couplings to the gauge theory survive and that 
the closed string interpretation of the UV/IR-terms naturally follows as 
a consequence of open-closed duality. The crucial feature that plays 
a role here is that the contributions from the massive modes cancel 
in this model.

This paper is organised as follows. In section 2 we give a brief review 
of open strings in background constant $B$-field and include the 
ingredients necessary for the computation of two point one loop open 
string amplitude. In section 3 we give a short review of strings on 
$C^2/Z_2$ orbifold and the massless spectrum of open strings ending on 
fractional $D_3$-brane localised at the fixed point. In section 4 we 
compute the two point function for one loop open strings and analyse it 
in the open and closed string channels. By taking the field theory 
limit, we show using open-closed string 
duality that the new IR divergent term from the nonplanar amplitude is 
exactly equal to the IR divergent contributions from massless closed 
string exchanges. We conclude this paper with discussions in section 5.

\noindent
{\it Conventions:} We will use capital letters $(M, N,...)$ to denote
general spacetime indices and small letters $(i,j, ...)$ for coordinates
along the $D$-brane. 


\section{Open superstring in Background $B$-field}

In this section we begin with some preliminaries of open string dynamics 
in background $B$-field that gives rise to noncommutative gauge theory 
on the $D_p$-brane in the Seiberg-Witten limit \cite{Seiberg}.
The world-sheet action for the bosons coupled to a constant $B$-field is 
given by,

\beqa
S_B&=&-\f{1}{4\pi\al}\int_{\Sigma}g_{MN}\pa_a X^M\pa^{a}X^N
+\f{1}{2}\int_{\pa\Sigma}B_{MN}X^M\pa_{\tau}X^N
\eeqa

The $B$-field is turned on only along the world volume directions of the 
$D_p$-brane, such
that, $B_{MN} \neq 0$ only for $M,N \leq p+1$ and $g_{MN}=0$ for $M \leq
p+1, N > p$.
The variation of the action gives the following boundary equations for
the world-sheet bosons,

\beqa
g_{MN}\pa_{\s}X^N+2\pi\al B_{MN}\pa_{\tau}X^N=0\mid_{\s=0,\pi}
\eeqa

The world-sheet propagator on the boundary of a disc satisfying this
boundary condition is given by,

\beqa
{\cal G}(y,y^{'})=-\al
G^{MN}\ln(y-y^{'})^2+\f{i}{2}\th^{MN}\ep(y-y^{'})
\eeqa

\noindent
where, $\ep(\Delta y)$ is $1$ for $\Delta y>0$ and $-1$ for
$\Delta y<0$. $G_{MN}$, $\th_{MN}$ are given by,

\beqal{gth}
G^{MN}&=&\left(\f{1}{g+2\pi\al B}g\f{1}{g-2\pi\al B}\right)^{MN}\non
G_{MN}&=&g_{MN}-(2\pi\al)^2(Bg^{-1}B)_{MN}\non
\th^{MN}&=&-(2\pi\al)^2\left(\f{1}{g+2\pi\al B}B\f{1}{g-2\pi\al
B}\right)^{MN}
\eeqa

The important point to note in the above relations is the difference 
between the open string metric, $G$ and the closed string metric $g$.
A nontrivial low energy theory results from the following limits,

\beqal{swl}
\al \sim \epsilon^{1/2} \rightarrow 0
\mbox{\hspace{0.1in};\hspace{0.1in}}
g_{ij} \sim \epsilon \rightarrow 0
\eeqa

\noindent
where, $i,j$ are the directions along the brane. This is the
Seiberg Witten (SW) limit which gives rise to noncommutative field
theory on the brane. The tree-level action for the low energy
effective field theory on the brane has the following form,

\beqa
S_{YM}=-\f{1}{g_{YM}^2}\int \sqrt{G}
G^{kk^{'}}G^{ll^{'}}\Tr(\hat{F}_{kl}*\hat{F}_{k^{'}l^{'}})
\eeqa

\noindent
where the $*$-product is defined by,

\beqa
f*g(x)=e^{\f{i}{2}\th^{ij}\pa_{i}^y\pa_{j}^z}f(y)g(z)\mid_{y=z=x}
\eeqa

\noindent
$\hat{F}_{kl}$ is the noncommutative field strength, and is
related to the ordinary field strength, $F_{kl}$ by,

\beqal{redeff}
\hat{F}_{kl}=F_{kl}+\th^{ij}(F_{ki}F_{lj}-A_i\pa_{j}F_{kl})+{\cal
O}(F^3)
\eeqa

\noindent
and,

\beqa
\hat{F}_{kl}=\pa_k\hat{A}_l-\pa_l\hat{A}_k-i\hat{A}_k*\hat{A}_l+
i\hat{A}_l*\hat{A}_k
\eeqa

We now include world-sheet fermions.
The action for the fermions coupled to $B$-field is given by,

\beqal{sf}
S_F=\f{i}{4\pi\al}\int_{\Sigma}g_{MN}\bps^{M}\rho^{\alpha}\pa_{\alpha}
\ps^{N}-\f{i}{4}\int_{\pa\Sigma}B_{MN}\bps^{N}\rho^0\ps^{M}
\eeqa

The full action including the bosons and the fermions with the bulk and 
the boundary terms are invariant under the following supersymmetry 
transformations,

\beqa
\de X^M&=&\bar{\ep}\psi^M\non
\de \psi^M&=&-i\rho^{\alp}\pa_{\alp}X^M\ep
\eeqa

We now write down the boundary equations by varying (\ref{sf})
with the following constraints,

\beqa
\de\ps^{M}_L=\de\ps^{M}_R\mid_{\s=\pi}
\mbox{\hspace{0.2in}and\hspace{0.2in}}
\de\ps^{M}_L=-(-1)^a\de\ps^{M}_R\mid_{\s=0}
\eeqa

where, $a=0,1$ gives the NS and the R sectors respectively
This gives the following boundary equations,

\beqal{bcf}
g_{MN}(\ps_L^N-\ps_R^N)+2\pi\al B_{MN}(\ps_L^N+\ps_R^N)&=&0
\mid_{\s=\pi}\\
g_{MN}(\ps_L^M+(-1)^a\ps_R^M) + 2\pi\al B_{MN}(\ps_L^N-(-1)^a\ps_R^N)
&=&0\mid_{\s=0}
\eeqa

\noindent
To write down the correlator for the fermions, first
define,

\beqa
\psi^M&=&\psi^M_L(\s,\tau) \mbox{\hspace{0.5in}} 0\leq\s\leq\pi\non 
&=&\left(\f{g-2\pi\al B}{g+2\pi\al B}\right)^M_N\psi_R^N
(2\pi-\s,\tau) \mbox{\hspace{0.5in}} \pi\leq\s \leq 2\pi
\eeqa
\noindent
This is the usual doubling trick that ensures the boundary conditions 
(\ref{bcf}). 
In the following 
section we would compute the two point function for the gauge field on 
the brane by inserting two vertex operators at the boundaries of the 
cylinder. Restricting ourselves to the directions along the brane, this 
vertex operator for the gauge field in the zero picture is given by,

\beqa
V(p,x,y)=\f{g_o}{(2\al)^{1/2}}\ep_j\left(i\pa_y X^j+4p.\Psi\Psi^j\right)
e^{ip.X}(x,y)
\eeqa

where $\Psi^i$ is given by,
\beqal{Psi}
\Psi^i(0,\tau)&=&\f{1}{2}\left(\psi^i_L(0,\tau)+(-1)^{a+1}\psi^i_R(0,\tau)\right)
=\left(\f{1}{g-2\pi\al B}g\right)^{i}_{j}\psi_L^j(0,\tau)\non
\Psi^i(\pi,\tau)&=&\f{1}{2}\left(\psi^i_L(\pi,\tau)+ 
\psi^i_R(\pi,\tau)\right)
=\left(\f{1}{g-2\pi\al B}g\right)^{i}_{j}\psi_L^j(\pi,\tau)
\eeqa

\noindent
Using (\ref{Psi}), the correlation function for $\Psi$ is given by,

\beqa
\expt{\Psi^i(w)\Psi^j(w^{'})}=G^{ij}{\cal G}\alpb (w-w^{'})
\eeqa

\noindent
where $G^{ij}$ is the open string metric defined in (\ref{gth}) and,
${\cal G}\alpb (w-w^{'})$ is given by \cite{narain},

\beqa
{\cal G}\alpb (w-w^{'})
=\f{\al}{4\pi}\f{\vth\alpb\left(\f{w-w^{'}}{2\pi},it\right)
\vth^{'}\oo (0,it)}{\vth\oo\left(\f{w-w^{'}}{2\pi},it\right)
\vth\alpb (0,it)}
\eeqa

\noindent
$\alp,\beta$ denotes the spin structures. $\alp=(0,1/2)$ are the NS 
and the R sectors and $\beta=(0,1/2)$ stands for the 
absence or the presence of the world-sheet fermion number $(-1)^F$ with 
$\psi$ being antiperiodic or periodic along the $\tau$ direction on the 
world-sheet. $w$, $w^{'}$ are located at the boundaries of the 
world-sheet for the open string which is a cylinder, i.e. at $\s=0,\pi$.

\subsection{Open strings on $C^2/Z_2$ orbifold and Fractional Branes}

An efficient and simple way to break ${\cal N}=4$ supersymmetry and  
obtain gauge theories with less supersymmetries is by orbifolding the 
background space. Specifically strings on $C^2/Z_2$ gives rise to ${\cal 
N}=2$ supersymmetric gauge theory on D3-branes with world volume 
directions transverse to the orbifolded planes.  
We will take the orbifolded directions to be $6,7,8,9$ with 
$Z_2=\{g_i\mid e,g\}$ such that $g^2=e$
The action of $g$ on these coordinates is given by,

\beqa
gX^I=-X^I \mbox{\hspace{0.5in} for \hspace{0.5in}} I=6,7,8,9
\eeqa

\noindent
In order to preserve world sheet supersymmetry we must consider the 
action of $Z_2$ on the fermionic partners, $\psi^I$. On a particular 
state the orbifold action is on the oscillators, $\psi^I_{-r}$ along 
with the Chan-Paton indices associated with it. Let us consider the 
massless bosonic states from the NS sector.

\beqa
g|i,j,\psi^I_{-1/2}>=\gamma_{ii^{'}}|i^{'},j^{'},\hat{g}\psi^I_{-1/2}>
\gamma^{-1}_{j^{'}j}
\eeqa

where $\gamma$ is a representation of $Z_2$
The spectrum is obtained by keeping the states that are invariant under 
the above action. To derive this it is easier to work in the basis 
where,

\beqa
\gamma=\s_3=\left(\begin{matrix} 1&0\\0&-1 \end{matrix}\right)
\eeqa

The action on the Chan-Paton indices can be thought of as,

\beqa
\gamma\left(\begin{matrix} 11&12\\21&22 \end{matrix}\right)\gamma^{-1}
=\left(\begin{matrix} 11&-12\\-21&22 \end{matrix}\right)
\eeqa

Thus the diagonal ones survive for the $Z_2$ action on the oscillators 
is $\hat{g}|\psi^I_{-1/2}>=|\psi^I_{-1/2}>$, i,e. for $I=2,3,4,5$ and 
the 
off diagonal ones are preserved for oscillators that are odd under the 
$Z_2$ action, $\hat{g}|\psi^I_{-1/2}>=-|\psi^I_{-1/2}>$, i,e. for 
$I=6,7,8,9$. The spectrum can thus be summarised as,

\beqa
A^I\rightarrow 
\left[\begin{matrix} A^{I}_1&A^{I}_2& \mbox{$2$ gauge fields}&I=2,3\\
                \phi^{I}_1&\phi^{I}_2&\mbox{$4$ real scalars}&I=4,5\\
             \Phi^{I}_1&\Phi^{I}_2&\mbox{$8$ real scalars}&I=6,7,8,9\\
\end{matrix}\right]
\eeqa

The above fields can be grouped into two vector multiplets and two 
hypermultiplets of ${\cal N}=2
$ with gauge group $U(1)\times 
U(1)$.
The beta function for the gauge couplings for this theory vanishes and 
the theory is conformally invariant. Now consider an irreducible 
representation $\gamma=\pm 1$. This acts 
trivially on the Chan-Paton indices. There is no image for the $D3$ 
brane. Following the above analysis we see that the spectrum consists of 
a single gauge field and two scalars completing the vector multiplet
of ${\cal N}=2$ with gauge group  $U(1)$. The beta function for this 
theory is nonzero.  With a constant 
$B$-field turned on along the world volume directions of the 
$D_3$-brane,$(0,1,2,3)$, the low energy dynamics on the brane will be 
described be noncommutative gauge theory in the Seiberg-Witten limit. 
In the following section we will study the
ultraviolet behaviour of this theory and see how the UV divergences have 
a natural interpretation in terms of IR divergences due to massless 
closed string modes as a result of open-closed string duality.

\section{Two point amplitude}

In this section we compute the two point function for the gauge fields 
on the brane. The necessary ingredients are given in section 2 and the 
appendix. The vacuum amplitude without any vertex operator insertion 
vanishes as a result of supersymmetry, i.e.

\beqal{zpt}
\det(g+2\pi\al B)\int^{\infty}_{0}\f{dt}{4t}(8\pi^2\al t)^{-2}
\sum_{(\alp,\beta,g_i)}Z\alpb_{g_i}=0 
\eeqa

\noindent
The factor of $\det(g+2\pi\al B)$ comes from the trace over the 
world sheet bosonic zero modes.
The sum is over the spin structures $(\alp,\beta)=(0,1/2)$ corresponding 
to the $NS-R$ sectors 
and the GSO projection and the orbifold projection. The elements 
$Z_{g_i} {\alpb}$ are computed in the 
appendix. Let us now compute the two point function. This is given by,

\beqal{2pt}
A(p,-p)&=&
\det(g+2\pi\al B)\int^{\infty}_{0}\f{dt}{4t}(8\pi^2\al t)^{-2}
\times \non &\times&\sum_{(\alp,\beta,g_i)}Z {\alpb}_{g_i}
\int_{0}^{2\pi t}dy\int_{0}^{2\pi t}dy^{'}
\expt{V(p,x,y)V(-p,x^{'},y^{'})}_{(\alp,\beta)}
\eeqa

\noindent
For the flat space, it is well known that amplitudes with less that four 
boson insertions vanish. However, in this model the two point amplitude 
survives. We will now compute this amplitude in the presence of 
background $B$-field. First note that the bosonic correlation function, 
$\expt{:\pa_yX^ie^{ip.X}::\pa_y^{'}X^ie^{-ip.X}:}$, does not 
contribute to 
the two point amplitude as it is independent of the spin structure. The 
two point function would involve the sum over the $Z_{g_i} \alpb$ which 
makes this contribution zero due to (\ref{zpt}). The nonzero 
part of the amplitude will be obtained 
from the fermionic part,

\beqa
\ep_k\ep_l\expt{:p.\Psi\Psi^ke^{ip.X}::
p.\Psi\Psi^le^{-ip.X}:}&=&\ep_k\ep_l p_ip_j
\left(G^{il}G^{jk}-G^{ij}G^{kl}\right)\times\\
&\times&{\cal G}^2\alpb (w-w^{'}) \expt{:e^{ip.X}::e^{-ip.X}:}\nn
\eeqa

\noindent
For the planar two point amplitude, both the vertex operators would be 
inserted at the same end of the cylinder (i.e. at $w=0+iy$ or $\pi+iy$). 
In this case, the sum in the two point amplitude reduces to,

\beqa
\sum_{(\alp,\beta,g_i)}Z\alpb_{g_i}{\cal G}^2\alpb(i\De y/2\pi)&=&
\sum_{(\alp,\beta)}Z\alpb_e{\cal G}^2\alpb(i\De y/2\pi)\non &+&
\sum_{(\alp,\beta)}Z\alpb_g{\cal G}^2\alpb(i\De y/2\pi)\nn
\eeqa
\beqa
&=&\f{4\pi^2}{\eta(it)^{6}\vth^2_1(i\De y/2\pi,it)}\sum_{(\alp,\beta)}
\vth^2(0,it)\alpb\vth^2\alpb(i\De y/2\pi,it)+ \\
&+& \f{16\pi^2}{\vth^{2}_{1}(i\De y/2\pi,it)
\vth^2_2(0,it)}\left[\vth^2_3(i\De
y/2\pi,it)\vth^{2}_{4}(0,it)-
\vth^2_4(i\De y/2\pi,it)\vth^{2}_{3}(0,it)\right]\nn
\eeqa

\noindent
where, $\De y=y-y^{'}$.
We have separated the total sum as the sum over the two $Z_2$ 
group 
actions. In writing this out we have used the following identity 

\beqa\eta(it)=\left[\f{\pa_{\nu}\vth_1(\nu,it)}{-2\pi}\right]^{1/3}_{\nu=0}
\eeqa

\noindent
Now, the first term vanishes due to the following identity

\beqa
\sum_{(\alp,\beta)}\vth\alpb (u)\vth\alpb (v)\vth\alpb (w)\vth\alpb (s)
=2\vth\oo (u_1)\vth\oo (v_1)\vth\oo (w_1)\vth\oo (s_1)
\eeqa

\noindent
where,
\beqa
u_1&=&\f{1}{2}(u+v+w+s) \mbox{\hspace{0.2in}} v_1=\f{1}{2}(u+v-w-s) 
\mbox{\hspace{0.2in}}\non w_1&=&\f{1}{2}(u-v+w-s) \mbox{\hspace{0.2in}}
s_1=\f{1}{2}(u-v-w+s)
\eeqa

\noindent
and noting that, $\vth\oo(0,it)=0$,
in the same way as the flat case that makes amplitudes with two 
vertex insertions vanish. The second term is a constant also due to,

\beqal{ident}
\vth^2_4(z,it)\vth^2_3(0,it)-\vth^2_3(z,it)\vth^2_4(0,it)
=\vth^2_1(z,it)\vth^2_2(0,it)
\eeqa

\noindent
For the nonplanar amplitude, which we are ultimately interested in, we 
need to put the two vertices at the two ends of the cylinder such that, 
$w=\pi+iy$ and $w^{'}=iy^{'}$. It can be seen that the fermionic part
of the correlator is constant and independent of $t$, same as the 
planar case following from the identity (\ref{ident}). The effect of 
nonplanarity 
and the regulation of the two point function due to the background 
$B$-field is encoded in the correlation functions for the exponentials. 
The two point function thus reduces to,

\beqa
A(p,-p)\sim \ep_k\ep_l p_ip_j
\left(G^{il}G^{jk}-G^{ij}G^{kl}\right)\int^{\infty}_{0}\f{dt}{4t}(8\pi^2\al 
t)^{-2}\int_{0}^{2\pi
t}dy dy^{'}\expt{e^{ip.X}e^{-ip.X}}
\eeqa

\noindent
The noncommutative gauge theory two point function is obtained in the 
limit
$t \rightarrow \infty$ and $\al \rightarrow 0$. The correlation function 
in this limit can be 
computed from the bosonic correlation functions \cite{oneloop,callan}.
We give below the function for the nonplanar case in this limit. 

\beqal{ti}
\expt{e^{ip.X}e^{-ip.X}}&=& \exp\left\{-p^2t \De x(\De x-1)-\f{1}{4t}
p_i(g^{-1}-G^{-1})^{ij}p_j\right\}\non &=&\exp\left\{-p^2t \De x(\De 
x-1)-\f{\tilde{p}^2}{4t}\right\}
\eeqa 

\noindent
where, $\tilde{p}=(\th p)$. We have redefined the world sheet coordinate 
as $\De x =\De y/(2\pi 
t)$ and have scaled $t \rightarrow t/(2\pi\al)$. We have also used the 
following relation in writing down the last expression.

\beqa
g^{-1}=G^{-1}-\f{(\th G \th)}{(2\pi\al)^2}
\eeqa

\begin{figure}[t]
\begin{center}
\begin{psfrags}
\psfrag{i}[][]{$\infty$}
\psfrag{z}[][]{$0$}
\psfrag{l}[][]{$\Lambda$}
\psfrag{Open}[][]{Open String Channel}
\psfrag{ls}[][]{$1/\Lambda^2\al$}
\psfrag{uv}[][]{UV}
\psfrag{p}[][]{$k$}
\psfrag{t}[][]{$t$}
\psfrag{a}[][]{(i)}
\psfrag{b}[][]{(ii)}
\psfrag{c}[][]{(iii)}
\psfrag{d}[][]{(iv)}
\psfrag{la}[][]{$1/\Lambda\al$}
\psfrag{las}[][]{$\Lambda^2\al$}
\psfrag{Closed}[][]{Closed String Channel}
\psfrag{ir}[][]{IR}
\psfrag{kpe}[][]{$\kpe$}
\psfrag{s}[][]{$s$}
\epsfig{file=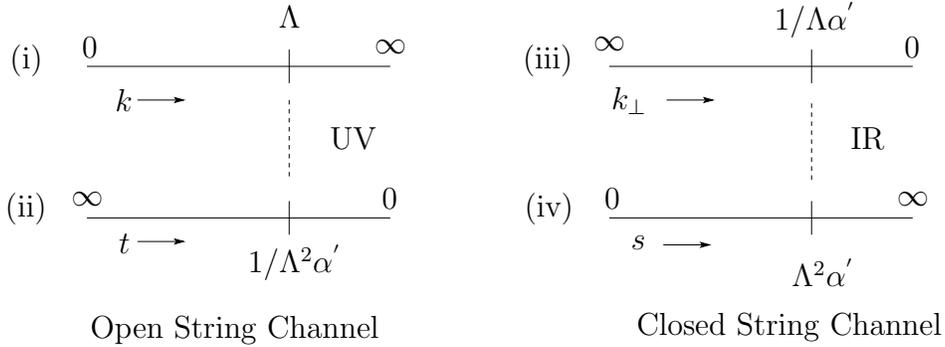, width= 12cm,angle=0}
\end{psfrags}
\vspace{ .1 in }
\caption{UV and IR regions in the open and closed string channels. (i) 
$k$ is the momentum in the gauge theory one loop diagram. (ii) $t$ is 
the modulus of the cylinder in the open string one loop diagram (iii) 
$\kpe$ is the transverse momentum of the closed string modes emitted 
from the brane (iv) $s=1/t$}
\end{center}
\end{figure}

\noindent
The first term in the exponential in (\ref{ti}) regulates the integral 
over $t$ in the infrared, for $p\neq 0$ and the 
second term regulates it in the ultraviolet that is 
usually observed in noncommutative field theories.
The $t \rightarrow \infty$ limit suppresses the contributions from all 
the open 
string massive modes. However as far as the field theory divergence 
is concerned they still come from the $t\rightarrow 0$ region. We can 
thus break the integral over $t$ into two intervals 
$1/\Lambda^2\al<t<\infty $ and $0<t<1/\Lambda^2\al$ (see Figure 1). The 
second one 
which is the source of the UV divergence is also the regime dominated by 
massless closed string exchanges. We now evaluate the two 
point function in this limit. First, the correlation function for the 
exponential in the $t \rightarrow 0$ limit is given by

\beqal{tz}
\expt{e^{ip.X}e^{-ip.X}}=\exp\left\{-\f{\al\pi}{2t}p_ig^{ij}p_j\right\}
\eeqa

\noindent
where $g^{ij}$ is the closed string metric. 
Modular transformation , ($t\rightarrow 1/t$) allows us 
to rewrite the 
one loop amplitude as the sum over closed string modes in a tree 
diagram.
In the limit $t\rightarrow 0$, the amplitude will be 
dominated by massless closed string modes.
In this model however, the effect of the massive modes in the loop 
cancel amongst themselves for any value of $t$. In the open string 
channel the $t \rightarrow 0$ limit would usually be contributed by the 
full tower of 
open string modes. However since we have seen that the effect of the 
massive string modes cancel anyhow for all values of $t$, the 
contribution to this limit from the open string modes comes only from 
the massless ones. The additional term in (\ref{ti}) as compared to 
(\ref{tz}) gives finite derivative corrections to the effective action. 
These would in general require the massive closed string states for its 
dual description. Without these derivative corrections, the 
contributions from the massless open string loop and the massless closed 
string tree are exactly equal.
The divergent ultraviolet 
behaviour of the massless open string modes can thus be captured by the 
the massless closed string modes that have momentum in the limit 
$[0,1/\Lambda\al]$. The amplitude can now be written as,

\beqa
A(p,-p)=V_4\det(g+2\pi\al B)\left(\f{g_o^2}{8\pi^2\al}\right)
\ep_k\ep_l p_ip_j
\left(G^{il}G^{jk}-G^{ij}G^{kl}\right)I(p)
\eeqa

\noindent
where,

\beqal{ip}
I(p)&=&\int ds s^{-1} 
\exp\left\{-\f{\al\pi s}{2}p_ig^{ij}p_j\right\}\non
&=&4\pi\int \f{d^2\kpe}{(2\pi)^2}\f{1}{\kpe^2+p_ig^{ij}p_j}
\eeqa

\noindent
The integral is written in terms of $s=1/t$ and in the last line we have 
rewritten it as an integral over $\kpe$, the momentum in the 
directions transverse to the brane for 
closed strings. The nonzero contribution to the two point amplitude 
in (\ref{2pt}) comes from the $\Tr_{NS}\left[gq^{L_0}\right]$ and 
$\Tr_{NS}\left[g(-1)^F q^{L_0}\right]$, that are evaluated
in (\ref{part}). These correspond to antiperiodic (NS-NS) and periodic 
(R-R) 
closed strings in the twisted sectors respectively. The fractional 
$D_3$-brane is localised at the fixed point of $C^2/Z_2$. Thus the 
twisted sector closed string states that couple to it are twisted in 
all the directions of the orbifold. 
These modes are localised at the fixed point and are free to move in the 
six directions transverse to the orbifold. This is the origin of the 
momentum integral (\ref{ip}) in two directions transverse to the 
$D$-brane. For the twisted sector the ground state energy for both the 
NS and the R sectors vanish. In the NS sector the massless modes 
come from $\psi^I_0$, $I=6,7,8,9$ oscillators which form a spinor 
representation of $SO(4)$. With the GSO and the orbifold projections, 
the closed string spectrum is given by, $2\times 2 =[0]+[2]$. The [0] 
and the self-dual [2] constitute the four massless scalars in the 
NS-NS 
sector. Similarly, in the R sector, the massless modes are given by 
$\psi^I_0$ for 
$I=2,3,4,5$. Thus giving a scalar and a two-form self-dual field in the 
closed string R-R sector. The couplings for the massless closed string 
states to the fractional $D_3$-brane have been worked out by various 
authors. See for example \cite{Bertolini:2001gq}.

As we are interested in seeing the ultraviolet effect of the 
open 
string channel as an infrared effect in the closed string channel, we 
must cut off the $s$ integral at the lower end at some value 
$\Lambda^2\al$ corresponding to the UV cutoff for the momentum 
of the massless closed strings in the directions transverse to the brane.

\beqa
I(p,\Lambda)\sim \int_{\Lambda^2\al}^{\infty}\f{ds}{s}e^{-\al p^2 s}\sim 
\int_0^{\infty}
d^2\kpe\f{e^{-(\kpe^2+p^2)\Lambda^2\al}}{\kpe^2+p^2}
\eeqa

The integral over $\kpe$ thus receives contributions upto $\kpe\sim 
{\cal O}(1/\Lambda\al)$, so that,

\beqal{final}
I(p,\Lambda)=4\pi^2\ln\left(\f{p_ig^{ij}p_j+1/(\Lambda\al)^2}
{p_ig^{ij}p_j}\right)
\eeqa

\noindent
This is the behaviour of the two point function for two gauge fields in 
${\cal N}=2$ theory. For the noncommutative theory it is regulated for 
$p\neq 0$. The fact that we are able to rewrite the gauge theory two 
point function as massless closed string tree-level exchanges is very 
specific to the ${\cal N}=2$ theory. The computations above show that 
the origin of this can be traced to open-closed string duality where the 
orbifold background cancels all contributions from the 
massive states as far as the UV singular terms are concerned. The 
background $B$-field in the SW limit only acts as a physical 
regulator.

\section{Discussions}

In the previous sections we have addressed the issue of open closed
duality in string theory in the presence of $B$-field. This duality 
lies
at the heart of duality between gravity and gauge theory as exemplified
by the AdS/CFT conjecture. In that situation the  gauge theory has N=4
supersymmetry and  is finite. The duality of the annulus diagram then reduces
to a trivial identity namely, 0=0. To get something non trivial one reduces
the amount of supersymmetry by orbifolding, but taking care to preserve enough
supersymmetry that there are no tachyons. In this case one  loop amplitudes are
divergent. One can compare divergences  in the closed and open string channels and
if one makes a suitable identification of the cutoffs one can show the
equality of amplitudes. What we have  done in this paper is to consider this
theory in the presence of $B$-fields  so that some of the amplitudes 
(non planar)
are actually finite and can be compared in an exact way. The $B$-field 
plays the role
of a regulator that preserves the duality. 

Once we turn on a $B$-field we also make contact with another 
phenomenon: UV/IR
mixing that is known to happen in non commutative field theories. This
acquires has a natural explanation when we consider this theory as the $\alpha '\rightarrow
0$ limit of a string theory a la Seiberg-Witten. The open
string loop UV region is reproduced by closed string trees with small
(i.e IR) momentum exchange. The $B$-field acts as a regulator for both 
amplitudes
but the regulation goes away as the external momentum goes to zero. It is not surprising
that the tree diagram diverges as the external momentum goes to zero, but by the duality
map this must also be true for the UV divergence of the gauge theory.

 In more detail we have studied a
noncommutative ${\cal N}=2$ gauge theory realised on a 
fractional $D_3$-brane localised at the fixed point of $C^2/Z_2$ 
orbifold. 

 The one loop two point open string amplitude gives the 
gauge theory two point amplitude in the $\alpha ' \rightarrow 0$ SW 
limit. This assumes there is no tachyon as is the case in the model studied here.
 We then see that the UV divergences of the gauge theory 
 comes from the $t\rightarrow 0$ end that is dominated by massless 
closed strings. In general  the massless closed 
string exchanges account for the UV contribution  due to all the open 
string modes. and similarly the dual description  of the gauge 
theory would thus require the contributions from all the massive closed 
string states as well. But in the supersymmetric case that is studied in 
this paper, the contributions from the massive modes cancel and hence 
the duality is between the finite number of massless states on both the open 
and the closed string sides. This is what is manifested through 
the equality of (\ref{final}) to the gauge theory amplitude, both ends 
being regulated by the presence of the $B$-field. 

To see the closed string coupling to the gauge theory, consider for the 
moment, the bosonic theory \cite{my}. The one loop open string amplitude 
is given 
by,

\beqal{op}
{\cal Z} \sim \int \f{dt}{t} (\al t)^{-\f{p+1}{2}} \eta(it)^{-(D-2)}
\exp(-C/\al t)
\eeqa

Here $C$ is the constant independent of $t$ and is a function of the 
$B$-field and external momenta. The expansion of the $\eta$-function in 
the $t \rightarrow \infty $ limit, in the the open string channel gives,

\beqal{opc}
{\cal Z}_{op} \sim \int \f{dt}{t} (\al t)^{-\f{p+1}{2}}
\left[e^{2\pi t}+(D-2)+O(e^{-2\pi t})\right]\exp(-C/\al t)
\eeqa
 
Let us set $\alpha ' t = T$
\beqal{1}
{\cal Z}_{op} \sim \int \f{dT}{T} (T)^{-\f{p+1}{2}}
\left[e^{2\pi {T\over \alpha '}}+(D-2)+O(e^{-2\pi {T\over \alpha '}})\right]\exp(-C/T)
\eeqa

The $O(1)$ term in the expansion corresponds to the massless open string 
modes in the loop. If we take the $\alpha ' \rightarrow 0$ 
limit the contribution of the massive modes drop out. If we ignore the tachyon
we get the massless mode contribution. In the supersymmetric case there is no tachyon.
However in the present case dropping the tachyon term makes an exact comparison
of the massless sectors of the two cases meaningless because the powers of $\alpha '$
cannot match. Nevertheless the comparison is instructive.  

The UV contribution of (\ref{opc}), as shown in 
Figure 1, comes from the region $0<t<1/\Lambda^2\al$. The UV 
divergences coming from this region is regulated by $C$. In the closed 
string channel we have,

\beqal{clc}
{\cal Z}_{cl} &\sim& \int ds (\al)^{-\f{p+1}{2}}s^{-l/2}
\left[e^{2\pi s}+(D-2)+O(e^{-2\pi s})\right]\exp(-Cs/\al )\non
&\sim& (\al)^{-\f{p+1}{2}} (\al)^{\f{l}{2}-1}\int 
d^l\kpe\f{1}{\kpe^2+C/\als}
\eeqa

The $\alpha '\rightarrow 0$ limit does not pull out the massless
sector (even if we ignore the tachyon) and this makes it clear
that in general all the massive closed string modes are required to 
reproduce the massless open string contribution.
But let us focus on the massless states of the closed string sector.
In the second expression of (\ref{clc}), we have kept only the 
contribution from the massless closed string mode.
This expression can be interpreted as the amplitude of emission and 
absorption of a closed string 
state from the $D_p$-brane with transverse momentum  $\kpe$, integrated 
over $0<\kpe<1/\Lambda\al$. The domain of the $\kpe$ integral 
corresponds to the UV region in the open string channel.   
$l=D-(p+1)$, is the number of transverse directions in which the 
closed string propagates. 
For $l\ne 2$ there is an extra factor of $(\al)^{\f{l}{2}-1}$
in (\ref{clc}) over (\ref{opc}), that makes the couplings of the 
individual closed string modes vanish when compared to the open string 
channel. It is only for the special case $l=2$ that the powers match.

In the supersymmetric case,
for the $C^2/Z_2$ orbifold, we have seen that the closed strings that 
contribute to the dual
description of the nonplanar divergences are from the twisted sectors.
They are free to move in 6 directions transverse to the orbifold. For 
the $D_3$-brane that is localised at the fixed point with world volume 
directions perpendicular to the orbifold, 
these closed string twisted states propagate in 
exactly two directions transverse to the brane. Thus in this case $l=2$ 
and from the above 
discussions this makes the power of $\al$ in the coupling of closed 
string 
with the gauge field strength same as that of the open string channel.

Although in general, the closed string couplings to the gauge field when 
the closed string modes are restricted to the massless ones do not 
give the same normalisation as the gauge theory, the massive closed 
string modes are expected to contribute so that the normalisations at 
both the ends are equal. This is guaranteed by open-closed string 
duality. For the $C^2/Z_2$ orbifold, since the massive states cancel, 
the finite number of closed string modes must give the same 
normalisation as the gauge theory two point function. This is the reason 
why we are able to see the IR behaviour of noncommutative ${\cal N}=2$ 
theory in terms of only the massless closed string modes in the twisted 
sectors.

The role played by the $B$-field is essentially that of a regulator that 
preserves the open closed duality. The 
fact that we see the UV divergence at the field theory level as IR 
divergence depends on the the special nature of this
 regulator that is dependent 
on the $B$-field and external momenta, thus giving rise to UV/IR mixing 
in noncommutative gauge theories. However the $B$-field does not affect 
the 
correspondence between the modes on the open and closed string sides 
that arise as a result of the 
world-sheet duality. The only modification of the partition function 
due to the $B$-field is the inclusion of a constant determinant (see 
appendix).
In conclusion, the IR divergences in 
noncommutative gauge theories that arise by integrating over high 
momentum modes in the loops can be seen as IR divergences due to closed 
string exchanges, as a result of open-closed string duality. The 
question of whether a finite or infinite number of closed string modes   
are necessary for the dual description depends on the commutative theory
without the  $B$-field.

At higher orders one expects the duality to be true for
the full string theory and not for the massless sectors. But in limits
such as in the AdS/CFT case one can expect a duality for the massless sectors. 
 It would be interesting to study the corresponding
AdS/CFT - like limit here. This is presumably some orbifolded version of the
AdS/CFT \cite{Kachru} with $B$-field \cite{nadscft}. 
\\
\\
\noindent
{\bf Acknowledgements :} One of us, S. Sarkar would like to
thank Bobby Ezhuthachan for many useful discussions.

\appendix
\section{Evaluation of Vacuum amplitude}
In this appendix we calculate the vacuum amplitude for the open strings 
with end points on a $D_3$-brane that is located at the fixed point of 
$C^2/Z_2$ orbifold. Let us first start with the bosonic part of the 
world-sheet action, 

\beqa
S_B=-\f{1}{4\pi\al}\int_{\Sigma}g_{MN}\pa_a X^M\pa^{a}X^N
+\f{1}{2}\int_{\pa\Sigma}B_{MN}X^M\pa_{\tau}X^N\\
\eeqa

The boundary condition for the world-sheet bosons from the above action 
is,

\beqal{bcb}
g_{MN}\pa_{\s}X^N+2\pi\al B_{MN}\pa_{\tau}X^N=0\mid_{\s=0,\pi}
\eeqa

In the Seiberg-Witten limit, $g_{ij}=\ep\eta_{ij}$ we choose the $B$
field along the brane to be of the form,

\beqal{bmatrix}
B=\f{\ep}{2\pi\al}
\left( \begin{array}{cccc}
0 & b_1 & 0 & 0\\
-b_1 & 0 & 0 &0 \\
0 & 0 & 0 & b_2 \\
0& 0& -b_2 & 0
\end{array} \right)
\eeqa

With the above form for the $B$-field, and defining,

\beqa
X_{(1)}^{\pm}=2^{-1/2}(X^0\pm X^1)
\mbox{\hspace{0.2in}and\hspace{0.2in}}
X_{(2)}^{\pm}=2^{-1/2}(X^2\pm iX^3)
\eeqa

the boundary condition (\ref{bcb}) can be rewritten as,

\beqa
\pa_{\s}X_{(1)}^{\pm}=\pm b_1\pa_{\tau}X_{(1)}^{\pm}\mid_{\s=0,\pi}
\mbox{\hspace{0.2in}and\hspace{0.2in}}
\pa_{\s}X_{(2)}^{\pm}=\pm ib_2\pa_{\tau}X_{(1)}^{\pm}\mid_{\s=0,\pi}
\eeqa

The mode expansions for the open string satisfying the above boundary 
conditions are given by,

\beqal{mb}
X^{\pm}_{(1)}&=&x^{\pm}_{(1)}+\f{2\al}{1-b_1^2}
(\tau\pm b_1\s)p^{\pm}_{(1)}
+i\sqrt{2\al}\sum_{n\neq 0}\f{a^{\pm}_{(1)n}}{n}e^{-i(n\tau \pm \nu_1)}
\cos(n\s \mp \nu_1)\non
X^{\pm}_{(2)}&=&x^{\pm}_{(2)}+\f{2\al}{1+b_2^2}
(\tau\pm ib_2\s)p^{\pm}_{(2)}
+i\sqrt{2\al}\sum_{n\neq 0}\f{a^{\pm}_{(2)n}}{n}e^{-i(n\tau \pm \nu_2)}
\cos(n\s \mp \nu_2)\non
\eeqa

\noindent
where we have defined,

\beqa
i\nu_1=\f{1}{2}\log\left(\f{1+b_1}{1-b_1}\right) \mbox{\hspace{0.5in}}
i\nu_2=\f{1}{2}\log\left(\f{1+ib_2}{1-ib_2}\right)
\eeqa

\noindent
The coefficients of the mode expansions (\ref{mb}) are fixed so as to 
satisfy,

\beqa
\left[X^{+}_{(1)}(\tau,\s),P^{-}_{(1)}(\tau,\s^{'})\right]=-2\pi\al\de(\s-\s^{'})
\eeqa

\noindent
and that the  zero modes and the other oscillators satisfy the usual 
commutation relations,

\beqa
\left[a^{+}_{(1)m},a^{-}_{(1)n}\right]=-m\de_{m+n}
\mbox{\hspace{0.5in}}
\left[a^{+}_{(2)m},a^{-}_{(2)n}\right]=m\de_{m+n}
\eeqa

\beqa
\left[x^{+}_{(1)},p^{-}_{(1)}\right]=-i \mbox{\hspace{0.5in}}
\left[x^{+}_{(2)},p^{-}_{(2)}\right]=i
\eeqa

There is no shift in the moding of the oscillators, the 
zero point energy and the spectrum is the same as the $B=0$ case. The 
situation is the 
same as that of a neutral string in electromagnetic background 
\cite{callan}. Note 
that the commutator for $X^{\pm}$ now does not vanish at the 
boundary, for example,

\beqa
\left[X^{+}_{(1)}(\tau,0),X^{-}_{(1)}(\tau,0)\right]=-2\pi i\al 
\f{b_1}{1-b_1^2}\non
\left[X^{+}_{(1)}(\tau,\pi),X^{-}_{(1)}(\tau,\pi)\right]=2\pi i\al 
\f{b_1}{1-b_1^2}
\eeqa

The zero mode for the energy momentum tensor can now be worked out and 
is given by,

\beqa
L_{(b)0}^{\parallel}=\f{2\al}{b_1^2-1}p^{+}_{(1)}p^{-}_{(1)}
+\f{2\al}{b_2^2+1}p^{+}_{(2)}p^{-}_{(2)}
-\sum_{n\neq 0}
\left[a^{+}_{(1)-n}a^{-}_{(1)}-a^{+}_{(2)-n}a^{-}_{(2)}
\right]
\eeqa

\noindent
Since the spectrum remains the same, the contribution to the vacuum 
amplitude from the bosonic modes is the same as the usual $B=0$ case 
except that there is a factor of 
$\sqrt{(b_i^2\pm 1)}$ which comes from the trace over the zero modes for 
each direction along the brane. From (\ref{bmatrix}) in 
the limit (\ref{swl}), $b_i \sim 1/\sqrt{\ep}$ for $B$ to be finite. 
With this,

\beqal{det}
\ep^2\prod_{i}^2(b_i^2 \pm 1) \rightarrow \det(g+2\pi\al B)
\eeqa

\noindent
Including contributions from all the directions,

\beqa
L_{(b)0}=L_{(b)0}^{\parallel}+L_{(b)0}^{\perp}+L_{(b)0}^{orb}-\f{5}{12}
\eeqa

\noindent
$\perp$ denotes the $4,5$ directions and $6,7,8,9$ are the orbifolded 
directions. Let us now compute the contributions from the world sheet 
fermions. The action is given by,

\beqa
S_F=\f{i}{4\pi\al}\int_{\Sigma}g_{MN}\bps^{M}\rho^{\alpha}\pa_{\alpha}
\ps^{N}-\f{i}{4}\int_{\pa\Sigma}B_{MN}\bps^{N}\rho^0\ps^{M}
\eeqa

We rewrite the boundary equations from (\ref{bcf}),

\beqal
g_{MN}(\ps_L^N-\ps_R^N)+2\pi\al B_{MN}(\ps_L^N+\ps_R^N)&=&0
\mid_{\s=\pi}\\
g_{MN}(\ps_L^M+(-1)^a\ps_R^M) + 2\pi\al B_{MN}(\ps_L^N-(-1)^a\ps_R^N)
&=&0\mid_{\s=0}
\eeqa

Now defining,

\beqa
\ps_{(1)R,L}^{\pm}=2^{-1/2}(\ps_{R,L}^0 \pm \ps_{R,L}^1)
\mbox{\hspace{0.2in}and\hspace{0.2in}}
\ps_{(2)R,L}^{\pm}=2^{-1/2}(\ps_{R,L}^2 \pm i\ps_{R,L}^3)
\eeqa

For the Ramond Sector $(a=1)$ with the constant $B$-field given by 
(\ref{bmatrix}),

\beqa
\ps_{(1)R}^{\pm}(1\pm b_1)=\ps_{(1)L}^{\pm}(1 \mp b_1)\mid_{\s=0,\pi}
\eeqa

Mode expansion,

\beqal{modeferm1}
\ps_{(1)L,R}^{\pm}=\sum_n d_{(1)n}^{\pm}\chi_{(1)L,R}^{\pm}(\s,\tau,n)
\eeqa

where,
\beqa
\chi_{(1)R}^{\pm}=\sqrt{2\al}\exp\{-in(\tau-\s)\mp\nu_1\}\\
\chi_{(1)L}^{\pm}=\sqrt{2\al}\exp\{-in(\tau+\s)\pm\nu_1\}
\eeqa

and
\beqa
\nu_1=\f{1}{2}\log\left(\f{1+b_1}{1-b_1}\right)=\tanh^{-1}b_1
\eeqa

The boundary condition for the other two directions are,

\beqa
\ps_{(2)R}^{\pm}(1\pm ib_2)=\ps_{(2)L}^{\pm}(1 \mp ib_2)\mid_{\s=0,\pi}
\eeqa

This gives the same mode expansion as (\ref{modeferm1}),

\beqal{modeferm2}
\ps_{(2)L,R}^{\pm}=\sum_n d_{(2)n}^{\pm}\chi_{(1)L,R}^{\pm}(\s,\tau,n)
\eeqa

\beqa
\chi_{(2)R}^{\pm}=\sqrt{2\al}\exp\{-in(\tau-\s)\mp\nu_2\}\\
\chi_{(2)L}^{\pm}=\sqrt{2\al}\exp\{-in(\tau+\s)\pm\nu_2\}
\eeqa

and
\beqa
\nu_2=\f{1}{2}\log\left(\f{1+ib_2}{1-ib_2}\right)=\tan^{-1}b_2
\eeqa

\noindent
Like the bosonic partners there is no shift in the frequencies. The 
oscillators are
integer moded as usual. For the Neveu-Schwarz sector, $(a=0)$, the 
relative sign between
$\ps_{R}^{\pm}$ and $\ps_{L}^{\pm}$ at the $\s=\pi$ end in eqn(6) can be 
brought about by the 
usual restriction on $n$ to only run over half integers in the mode
expansions (\ref{modeferm1},\ref{modeferm2}).
The oscillators satisfy the standard anticommutation relations,

\beqa
\{d_{(1)n}^{+},d_{(1)m}^{-}\}=-\de_{m+n}
\mbox{\hspace{0.2in};\hspace{0.2in}}
\{d_{(2)n}^{+},d_{(2)m}^{-}\}=\de_{m+n}
\eeqa

\noindent
The zero mode for the energy momentum tensor for the fermions along 
the brane can be written as,

\beqa
L_{(f)0}^{\parallel}=\sum_n 
n\left[d_{(2)-n}^{-}d_{(2)n}^{+}-d_{(1)-n}^{-}d_{(1)n}^{+}\right] 
\eeqa

\noindent
For all the fermions including the contributions from the other 
directions we have,

\beqa
L_{(f)0}=L_{(f)0}^{\parallel}+L_{(f)0}^{\perp}+L_{(f)0}^{orb} +c_f(a)
\eeqa

\noindent
where $L_{(f)0}^{\perp}$ and $L_{(f)0}^{orb}$ have the usual 
representation in terms of oscillators.

\beqa
c_f(1)=\f{5}{12} \mbox{\hspace{0.2in};\hspace{0.2in}} c_f(0)=-\f{5}{24}
\eeqa

\noindent
We now compute the vacuum amplitude including the contributions from the 
ghosts. This is given by,

\beqa
Z_C=V_4\det(g+2\pi\al B)\int_0^{\infty}
\f{dt}{t}(8\pi^2\al t)^{-2}\Tr_{NS-R}\left[\left(\f{1+g}{2}\right)
\left(\f{1+(-1)^F}{2}\right)q^{L_0}\right]
\eeqa

\noindent
The origin of the $\det(g+2\pi\al B)$ term is given in (\ref{det}) and 
$V_4$ is the volume of the $D_3$-brane and $q=e^{-2\pi t}$.
The trace is summed over the spin structures with the orbifold 
projection. The required traces are listed below in terms of the {\it 
Theta Functions}, $\vth_i(\nu,it)$ (see for example \cite{polchinski}).

\beqal{part}
Z\zz_{e}(it)&=&\Tr_{NS}\left[q^{L_0}\right]= 
\eta(it)^{-12}\vth^4_{3}(0,it)\\
Z\zo_e(it)&=&\Tr_{NS}\left[(-1)^F q^{L_0}\right]=
-\eta(it)^{-12}\vth^4_{4}(0,it)\non
Z\oz_{e}(it)&=&\Tr_{R}\left[q^{L_0}\right]=
-\eta(it)^{-12}\vth^4_{2}(0,it) \non
Z\oo_e(it)&=&\Tr_{R}\left[(-1)^F q^{L_0}\right]=0\non
Z\zz_{g}(it)&=&\Tr_{NS}\left[gq^{L_0}\right]= 
4\eta(it)^{-6}\vth^2_{3}(0,it)\vth^2_{4}(0,it)\vth^{-2}_{2}(0,it)\non
Z\zo_g(it)&=&\Tr_{NS}\left[g(-1)^F q^{L_0}\right]=
-4\eta(it)^{-6}\vth^2_{3}(0,it)\vth^2_{4}(0,it)\vth^{-2}_{2}(0,it)\non
Z\oz_{g}(it)&=&\Tr_{R}\left[gq^{L_0}\right]=0 \non
Z\oo_g(it)&=&\Tr_{R}\left[g(-1)^F q^{L_0}\right]=0\non
\eeqa

Recalling,

\beqa
\vth^4_{3}(0,it)-\vth^4_{4}(0,it)-\vth^4_{2}(0,it)=0
\eeqa

and noting that,

\beqa
Z\zz_g(it)=-Z\zo_g(it) 
\eeqa

the vacuum amplitude vanishes. This is as a result of supersymmetry.

\end{document}